\documentclass[final,leqno,onefignum,onetabnum]{siamltex1213}%

\usepackage{amssymb,amsmath}
\usepackage{color}
\usepackage{multirow}%

\newtheorem{rem}{Remark}

\title{Predicting Triadic Closure in Networks Using Communicability Distance
Functions}

\author{Ernesto Estrada\footnotemark[2] 
\and Francesca Arrigo\footnotemark[3]}

\begin{document}
\maketitle

\footnotetext[2]{Department of Mathematics and Statistics, University of Strathclyde,
26 Richmond Street, Glasgow G1 1XQ, U.K. (\email ernesto.estrada@strath.ac.uk). } 
\footnotetext[3]{Department of Science and High Technology, 
University of Insubria, Como 22100, Italy (\email 
francesca.arrigo@uninsubria.it).}


\begin{abstract}
We propose a communication-driven mechanism for predicting triadic
closure in complex networks. It is mathematically formulated on the
basis of communicability distance functions that account for the quality
of communication between nodes in the network. We study $25$ real-world
networks and show that the proposed method predicts correctly $20\%$
of triadic closures in these networks, in contrast to the $7.6\%$
predicted by a random mechanism. We also show that the communication-driven
method outperforms the random mechanism in explaining the clustering
coefficient, average path length, and average communicability. The
new method also displays some interesting features with regards to optimizing 
communication in networks. 
\end{abstract}


\pagestyle{myheadings}
\thispagestyle{plain}
\markboth{{\sc Ernesto Estrada and Francesca Arrigo}}{{Predicting triadic closure}}


\section{Introduction}

Complex networks are ubiquitous in many real-world scenarios, ranging
from the biomolecular---those representing gene transcription, protein
interactions, and metabolic reactions---to the social and infrastructural
organization of modern society \cite{LucianoReview,Estrada2011book,NewmanReview}.
Mathematically, these networks are represented by graphs, where the
nodes represent the entities of the system and the edges represent
the ``relations'' among those entities. The accumulation of a mountain 
of empirical
evidence has left little doubt that in general 
real-world networks are very different from their random counterparts
in many structural and functional aspects \cite{Estrada2011book}.
In particular, it is well-documented that real-world networks
are significantly more ``clustered'' than one would expect
from a random wiring of nodes \cite{NewmanReview}. The degree of
``clustering'' is usually quantified in network theory through the
use of the \textit{clustering coefficient} (see \cite{WattsStrogatz}). This 
accounts for the ratio of the number of triangles to the number
of open triads, i.e. subgraphs of the type $i-j-k$. The fact that
triangles are abundant in real-world networks has long been appreciated---for 
example, in 1922 where Simmel \cite{Simmel} theorized that people with 
common
friends are more likely to create friendships. This ``\textit{friendship
transitivity}'' definitively implies a social mechanism for triadic
closure in social networks which may then be applied to explain the
evolution of triangle closures \cite{krackhardt_Handcock}. This Simmelian
principle of triadic closure due to friendship transitivity assumes
that individuals can benefit from cooperative relations, and this may induce
individuals to choose new acquaintances from among their friends'
friends.

The high degree of transitivity is not a unique feature of social
networks; indeed, it is a common characteristic of many other types
of networks such as biomolecular, cellular, ecological, infrastructural,
and technological (see \cite{Estrada2011book} and references therein).
It is natural to assume that analogous cooperative principles to the
one proposed by Simmel for social networks could be applied to find
mechanisms that explain triadic closure in these other types of networks.
Although intuitive, this simple idea has some fundamental drawbacks.
First, it is not always true that pairs of nodes benefit from cooperative
relations, and therefore the Simmelian principle is useless in such
situations. Secondly, it is evident that not every pair of nodes separated
by two edges participates in a triangle in a real-world network. Thus,
some kind of selective process has been taking place, closing some
of the triads in a network and leaving many others open.

The goal of this paper is to propose a general mechanism to account
for such selective process of triadic closure in networks. We propose
a strategy for predicting triadic closure based on the idea that triadic
closure is a communication-driven process. This paradigm is formulated
on the basis of communicability distance functions that account for
the quality of communication between pairs of nodes using a mechanism 
accounting for both local and long-range interactions. We start with
an overview of related work. All the mathematical concepts we
use are introduced in Section \ref{sec:MP} in order to make the paper 
self-contained. Sections
\ref{sec:CD} and \ref{sec:PM} are devoted to the introduction of
the new method for predicting triadic closure. We finish with a
presentation and discussion of the results.

\section{Related Work}

\label{sec:RW}

Triadic closure, loosely defined as the process in which an edge is
added to a triad to form a triangle, has long been considered as a
fundamental mechanism of social networks' evolution. The theoretical
basis of this mechanism is due to Simmel \cite{Simmel} and one of the 
pioneering studies to use this principle to predict triadic
closure in social networks was published by Krackhardt and Handcock
\cite{krackhardt_Handcock}.

When considering undirected networks, the main focus of triadic closure
models has been to create simple mechanisms that provide insight into 
how (social) networks grow and generate their main topological characteristics.
A simple model of network growth based on triadic closure has been
proposed by Bianconi et al. \cite{TriadClosureCommuties}. They show
that the evolution of networks based on such simple mechanisms ``\textit{naturally
leads to the emergence of community structure, together with fat-tailed
distributions of node degree and high clustering coefficients}''.
Similar results by Klimek and Thurner \cite{KlimekThurner} suggest that triadic
closure can be identified as one of the fundamental dynamical principles in
social multiplex network formation. These two works use triadic
closure mechanisms based on the random selection of the nodes which
will be involved in the triangles.

In the case of directed graphs an exhaustive computational analysis
was performed by Leskovec et al.~\cite{Leskovec}. They consider
several strategies to model how a node $u$ selects a node $v$, two
steps from it, to form a triangle. The basic strategy is for $u$ to
select randomly a node $v$ from all the nodes at distance two.
An alternative strategy is to assume that $u$ first selects
a neighbor node $w$ according to some mechanism, and then $w$ selects
a neighbor $v$ according to some (possibly different) mechanism.
The edge $(u,v)$ is then formed and the triangle $\triangle_{uwv}$
is closed. The selection of a neighbor $w$ for $u$ (or $v$ for
$w$) has been carried out using the following techniques: $(i)$
uniformly at random; $(ii)$ proportional to degree of $w$ raised
to a power; $(iii)$ proportional to the number of friends that $u$
and $w$ have in common; $(iv)$ proportional to the time passed since
$w$ last created an edge raised to a power; $(v)$ proportional to
the product of the number of common friends of $u$ and $w$ multiplied by the 
last activity time, all raised to a power.

The quantitative predictions made by Leskovec et al. are summarised in
Table 1, where we report the percentage of correct prediction of triadic
closure in four networks, using a random-random selection of the nodes,
and the best percentage of improvement over the log-likelihood of
picking a random node two hops away (baseline) reported by the authors.
The network LinkedIn is the only one of the four which is undirected.

\begin{table}
\centering
\footnotesize%
\begin{tabular}{|c|c|c|}
\hline 
\multirow{2}{3cm}{Network} & \multicolumn{2}{c|}{ \% correct triadic closure}\\
\cline{2-3} 
 & random-random & best
\tabularnewline
\hline 
\hline 
Flickr    &  13.6 & 16.9 \\
\hline 
Delicious &  11.7 & 18.2 \\
\hline 
Answers   &   6.8 & 16.4 \\
\hline 
Linkedin  &   16.0 & 21.4\\
\hline 
\end{tabular}
\caption{Illustration of the percentage of correct prediction
of triadic closure in online social networks by the random-random
selection of nodes and the best of all predictions made by Leskovec
et al. \cite{Leskovec}.}
\end{table}

In a more recent paper, Lou et al.~\cite{Lou13} have developed a
method that adds sociological information to the network structure
in order to predict triadic closure in a Twitter network. Their approach
uses information about $(i)$ geographic distance, i.e. whether users
have a higher probability of following each other when they are located
in the same region; $(ii)$ homophily, i.e. whether similar users
tend to follow each other; $(iii)$ implicit network, i.e. how
the following network on Twitter correlates with other implicit
networks, such as the retweet and reply network; $(iv)$ social
balance, i.e. whether the reciprocal relationship network on Twitter
satisfies social balance theory and to what extent. When this
non-topological information is added, the developed method outperforms
other structure-only approaches in the prediction of triadic closures.
A similar approach, which uses demographic information instead, has
been developed by Huang et al.~\cite{Huang14}. They have used a
large microblogging network as the source of their study, which reveals
how user demographics and network topology influence the process of
triadic closure. Their experimental results on the microblogging data
show the efficiency of the proposed model for the prediction of triadic
closure formation.

Here we will not account for extra-topological information, i.e.,
information apart from that provided by the topological structure
of the network. Thus, our current work is more in the spirit of that
of Leskovec et al. \cite{Leskovec} with the difference that the networks
we study are undirected.


\section{Mathematical Preliminaries}

\label{sec:MP}

A \textit{graph} $\Gamma=(V,E)$ is defined by a set of $n$
nodes (vertices) $V$ and a set of $m$ edges $E=\{(u,v)|u,v\in V\}$
between the nodes. An edge is said to be \textit{incident} to a vertex
$u$ if there exists a node $v\neq u$ such that either $(u,v)\in E$
or $(v,u)\in E$. The \textit{degree} of a vertex $u$, denoted by
$d_{u}$, is the number of edges incident to $u$ in $\Gamma$. The
graph is said to be \textit{undirected} if the edges are formed by
unordered pairs of vertices. A \textit{walk} of length $k$ in $\Gamma$
is a set of nodes $u_{1},u_{2},\ldots,u_{k},u_{k+1}$ such that for
all $1\leq l\leq k$, $(u_{l},u_{l+1})\in E$. A \textit{closed walk}
is a walk for which $u_{1}=u_{k+1}$. A \textit{path} is a walk with
no repeated nodes. A closed walk of length 3 is called a \textit{triangle}.
We will call \textit{triad} every triplet of nodes $u$, $v$, and
$w$ such that $(u,v),(v,w)\in E$ but $(u,w)\not\in E$. Hence a triad
is a triangle missing one edge. We shall call this
missing edge a \textit{potential edge}. A graph is \textit{connected}
if there is a path joining $u$ and $v$ for every $u,v\in V$. A
graph with unweighted edges, no edges from a node to itself, and no
multiple edges is said to be \textit{simple}.

Let $A=\left(a_{uv}\right)\in\mathbb{R}^{n\times n}$ be the \textit{adjacency
matrix} of the graph. It is worth noting that for undirected, simple,
and connected networks the associated adjacency matrix is symmetric,
binary, hollow (i.e., has zeros on the main diagonal), and irreducible
(see \cite{HornJohnson}) and its entries are: 
\[
a_{uv}=\left\{ \begin{array}{ll}
1 & \mbox{if }(u,v)\in E\\
0 & \mbox{otherwise}
\end{array}\right.\qquad\forall u,v\in V.
\]

It is possible to define several distance measures on networks. The
most common is the \textit{shortest-path} (or \textit{geodesic}) \textit{distance}
between two nodes $u,v\in V$, which is defined as the length of the
shortest path connecting these nodes. We will write $d(u,v)$ to denote
the geodesic distance between $u$ and $v$. and hence the \textit{average path 
length} \cite{Estrada2011book,NewmanReview}, the average
of the shortest path distances in the graph, is given by
\[
\overline{\ell}=\frac{1}{2m}\sum_{u,v\in V}d(u,v).
\]
Another useful measure for characterizing the structure of networks
is the so-called \textit{local clustering coefficient} of a node $u$ 
\cite{WattsStrogatz},
which quantifies the degree of transitivity of local relations in
a network and is defined as
\[
C_{u}=\frac{2t_{u}}{d_{u}(d_{u}-1)},
\]
where $t_{u}$ is the the number of triangles in which node $u$ participates.
Taking the mean of these values as $u$ varies among all the nodes
in $\Gamma$ gives the \textit{clustering coefficient} of the network,
\[
\overline{C}=\frac{1}{n}\sum_{u=1}^{n}C_{u}.
\]

An important quantity to be considered when studying communication
processes in networks is the \textit{communicability function} \cite{Communicability,Estrada Hatano Benzi,Estrada Higham},
which is defined as
\[
G_{uv}=\left(e^{A}\right)_{uv}=\sum_{k=0}^{\infty}\frac{\left(A^{k}\right)_{uv}}{k!}=\sum_{k=1}^{n}e^{\lambda_{k}}\mathbf{q}_{k}(u)\mathbf{q}_{k}(v),\qquad\forall u,v\in V,
\]
where $A=Q\Lambda Q^{T}$ is the spectral decomposition of the adjacency
matrix (see \cite{HornJohnson}), with $\Lambda$ a diagonal matrix
containing the eigenvalues of $A$ and $Q=[\mathbf{q}_{1},\ldots,\mathbf{q}_{n}]$
an orthogonal matrix containing the associated eigenvectors.

Communicability counts the total number of walks starting at node
$u$ and ending at node $v$, weighting their length by a factor $\frac{1}{k!}$,
thus considering shorter walks more influential than longer ones.
The $G_{uu}$ terms of the communicability function, which are usually
called \textit{subgraph centralities} of the nodes, characterize the
degree of participation of a node in all subgraphs of the network,
giving more weight to the smallest ones. Here we will use the \textit{average
communicability} as a way to characterize the quality of the communication
taking place in the network as a whole: 
\[
\overline{G}=\frac{1}{n(n-1)}\sum_{u\neq v}G_{uv}
\]

The communicability function can be used to 
quantify the quality of communication between
nodes in a network. When two nodes $u$ and $v$ are exchanging information,
the quality of their communication depends on two factors:
how much information departing from a source node reaches its target
($G_{uv}$), and how much information departing from the node
returns to it without ending at its destination ($G_{uu}$). That
is, the quality of communication increases with the amount of information
that departs from the originator and arrives at its destination,
and decreases with the amount of information which is wasted due
to the fact that the information returns to its source without being
delivered to the target. In \cite{CommDist} the \textit{communicability
distance} is defined as
\begin{equation}
\xi_{uv}=\sqrt{G_{uu}+G_{vv}-2G_{uv}}.\label{eq:xi}
\end{equation}
It is a  Euclidean distance between the nodes
$u$ and $v$ in $\Gamma$ (see \cite{CommDist,CommDistAppl}). From
its definition, it is clear that $\xi_{uv}$ characterizes the quality
of the communication taking place between nodes $u$ and $v$.


\section{Communicability Distances and Triad Closure}

\label{sec:CD}

We start by considering the square of the communicability distance 
defined in \eqref{eq:xi} for a pair of nodes $uv$ in a connected
graph. This distance characterizes  communication quality between
nodes $u$ and $v$ by assuming that the information departing
from node $u$ travels to node $v$ (and viceversa) by taking a series of one-hop
steps between the nodes in any of the walks that connect
them. From \eqref{eq:xi}, it is clear that the smaller
the value of $\xi_{uv}^{2}$, the better nodes $uv$ are at exchanging
information. The communicability distance is dependent on $e^{A}$, where $A$ is 
the adjacency matrix of a simple graph.
If we consider $u$ and $v$ such that $a_{uv}=1$, then we are assuming
that these two nodes are attracted to each other. If instead we
were to consider that these nodes repel each other, we would use $e^{-A}$.

If the (squared) communicability distance between two pairs of nodes $uv$ and 
$pq$ satisfies $\xi_{uv}^{2}<\xi_{pq}^{2}$ 
then we say that the attraction between the pair $uv$ is stronger
than that of the pair $pq$ in the corresponding network.

Now, consider a triad $u,w,v$, where $(u,w)\in E$, $(w,v)\in E$ but 
$(u,v)\notin E$. Because $a_{uw}=1$ and $a_{wv}=1$
we can infer that there are attractive ``\textit{forces}'' between
$u$ and $w$ and between $w$ and $v$. A simple metaphoric way to represent such 
attractive forces between pairs of nodes is to suppose
that they have opposite charges which attract to each other. For instance,
we can consider either of the following schemes for the previous example:
$u^{+}-w^{-}-v^{+}$, $u^{-}-w^{+}-v^{-}$. Notice that considering
a particle spin, as is usually done in sociophysical models of opinion
dynamics, also works here as an appropriate metaphor (see for
instance \cite{Sen}). Observe that there are two types of interactions
between the nodes $u$ and $v$. First, due to the attractions between
$u$ and $w$ and $w$ and $v$, the node $v$ `feels' an attractive
force from $u$, which is transmitted through the edges of the network.
On the other hand, due to the fact that both $u$ and $v$ have the same charge,
they experience some repulsion from each other, which takes place
in a `through-space' fashion (which we will clarify later). We can expect
the link $(u,v)$ to be created if the through-edge
attractive force between the nodes $u$ and $v$ is larger than the
through-space repulsive force between them.

In order to understand the nature of the interactions described in
the previous paragraph we consider a molecular system as a model example.
In this case there is a communication between pairs of atoms which
occurs through the covalent bonds of the molecule. This kind of interaction
takes place through the edges (covalent bonds) and is analogous
to the attractive forces we have previously described. Hereafter we
will refer to this interaction as the \textit{Through-Edges Communicability
(TEC)}. If two non-covalently bonded atoms are close in space, they
can interact with each other through non-covalent interactions, for
example, by hydrophobic, polarity or electrostatic forces. These interactions
are analogous to our through-space repulsion and we will refer to them as 
\textit{direct Long-Range Communicability (LRC)}.
In a social network, TEC is present when information is transmitted
from one individual to another in the network by using the social
ties that define the edges of the graph. On the other hand, LRC
is realized by the direct influence of an individual to another through any 
source of social
signalling.

Note that although the shortest path distance between every
pair of nodes in a triangle equals one, every pair of nodes in it
is connected by a pair of adjacent edges through the third vertex.
A natural way to account for all the pairs of nodes connected
by pairs of adjacent edges is to consider the number of walks of length
two between the pairs of nodes. We can then transform a graph accordingly. Let 
$\Gamma=(V,E)$ be a simple and connected graph and let $W_{2}(\Gamma)=(V,E')$ 
be the graph with the same set of nodes
as $\Gamma$ but whose edges are weighted by
the number of walks of length two between every pair of (not necessarily
distinct) nodes in $\Gamma$. More precisely, if $\mu_{2}(u,v)$ is
the number of walks of length two between nodes $i$ and $j$,
then the adjacency matrix $\tilde{A}$ of $W_{2}\left(\Gamma\right)$
is
\[
(\tilde{a})_{uv}=\left\{ \begin{array}{ll}
\mu_{2}(u,v) & u\neq v\\
\mu_{2}(u,u)=d_{u} & u=v.
\end{array}\right.
\]

\begin{rem} Clearly $\tilde{A}=A^{2}$ and so we do not need to explicitly 
construct
the graph $W_{2}\left(\Gamma\right)$, since we can simply work with
the square of the adjacency matrix of the graph $\Gamma$. \end{rem}

Note that two nodes are connected in $W_{2}\left(\Gamma\right)$
if they have the same charge and so connected nodes
in $W_{2}\left(\Gamma\right)$ repel each other. Consequently the
repulsive communicability between a given pair of nodes in $\Gamma$
is given by $\tilde{G}_{uv}=(e^{-\tilde{A}})_{uv}=(e^{-A^{2}})_{uv}$.

A communicability distance based on $\tilde{G}_{uv}$ accounts for
the quality of LRC between pairs of nodes separated by
two adjacent edges, i.e., pairs of nodes feeling mutual repulsion
in $\Gamma$. We can define a communicability distance 
function by
\begin{equation}
\eta_{uv}=\sqrt{\tilde{G}_{uu}+\tilde{G}_{vv}-2\tilde{G}_{uv}}\label{eq:eta}
\end{equation}
A large value of $\eta_{uv}$ indicates that there
is a weak repulsion between nodes $u$ and $v$.
The proof that $\eta_{uv}$ is a Euclidean distance between the nodes
$u$ and $v$ follows the same lines as in \cite{CommDist,CommDistAppl}
and is omitted.
\begin{rem} The graph $W_{2}\left(\Gamma\right)$
is not always connected and so the function $\eta_{uv}$ is defined
only for pairs of nodes which are in the same connected component
of the graph. Elsewhere $\eta_{uv}$ is set to infinity. \end{rem}

\begin{figure}
\centering 
\includegraphics[width=0.8\textwidth]{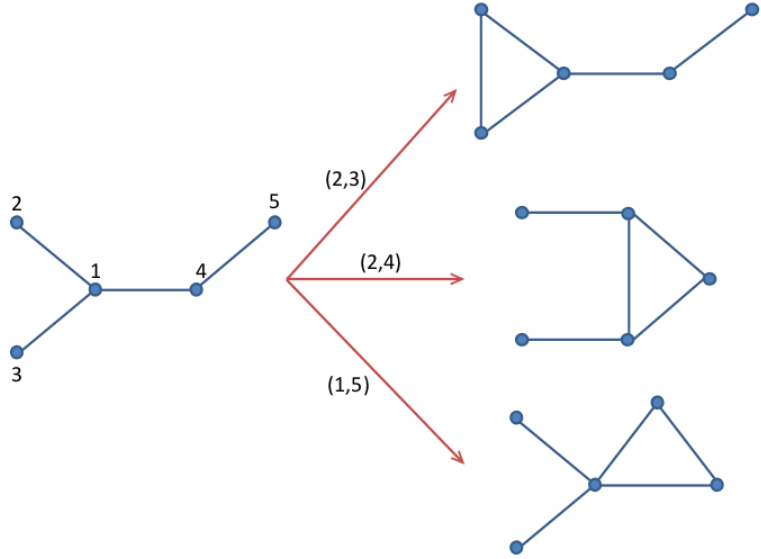} 
\caption{Example of evolution of a tree after one edge is added to close a
triangle.}
\label{transformation}
\end{figure}

Before continuing,  consider the following example. The tree
illustrated on the left in Figure 1 can be transformed by adding an edge 
which
closes any of the three nonequivalent existing triads of the graph,
i.e. by adding the edge $(2,3)$, $(2,4)$ or $(1,5)$. The resulting
unicyclic graphs are illustrated on the right of Figure
1. In Table \ref{Table of transformations} we report the values of
$\xi_{uv}^{2}$ and $\eta_{uv}^{2}$ for each of the three triads. Now assume 
that we have information indicating
that the process giving rise to the closure of the $1,2,3$-triad
is favored over the other two. We cannot known \textit{a priori} for
any particular system how the attractive and repulsive forces scale. In real 
physical systems such terms are scaled by minimizing
the global energy of the system. Here we simply consider the weighted
difference between the two terms, $\alpha\xi_{uv}^{2}-\beta\eta_{uv}^{2}$.
We will propose a method to determine the values of the empirical
parameter $\alpha$ and $\beta$ in a given network a little later. For this
example it can be verified that, for instance, $\xi_{uv}^{2}-1.5\eta_{uv}^{2}$
produces a negative value only for the pair $(2,3)$ (see Table \ref{Table of transformations}).
This weighted difference between $\xi_{uv}^{2}$ and $\eta_{uv}^{2}$
corresponds to the case in which the attractive forces between the
corresponding nodes outweight the magnitude of the repulsive ones.
As noted previously, a large value of $\eta_{uv}^{2}$ indicates
a small repulsion between the corresponding nodes, and here we have
multiplied $\eta_{uv}^{2}$ by a coefficient $\beta>1$, which further reduces 
the repulsive forces.

Now suppose instead that we have information indicating that the process
giving rise to the closure of the $1,4,5$-triad is favored over the
other two. In this case it can be verified (see Table \ref{Table of transformations})
that the weighted difference $-0.5\xi_{uv}^{2}+1.5\eta_{uv}^{2}$
is negative only for the pair $(1,5)$. Here, we have considered that
the attractive forces between the nodes make a negative contribution
to the creation of an edge closing the triad. This may correspond
to the situation in which the links $(u,w)$ and $(w,v)$ are both
very weak, i.e. friendship ties between the corresponding
individuals are not too strong. We further weaken those relations by 
multiplying
$\xi_{uv}^{2}$ by a coefficient $\alpha<0$. At the same time, by multiplying 
$\eta^2_{uv}$ by a coefficient $\beta<0$ we
have assumed that the repulsive factor does not play a major role
in determining whether the new edge is created or not. Indeed, in this
way $\eta_{uv}^{2}$ is transformed into an attraction term. In the
charged-particles analogy this  corresponds to a situation in which
the charges between the corresponding nodes are very weak and there is
no repulsion between those nodes separated by two adjacent edges.

Finally, suppose that we have information indicating that the process
giving rise to the closure of the $1,2,4$-triad is favored over the
other two. In this case it can be verified (see Table \ref{Table of transformations})
that the weighted difference $\xi_{uv}^{2}+\eta_{uv}^{2}$ reaches
the smallest value for $(2,4)$. The values of the weighted differences
for the three triad closure processes are positive, but the one corresponding
to the closure of he $1,2,4$-triad is the lowest among the three.
In this case, triadic closure is dominated by attractive forces
only. The term $\alpha\xi_{uv}^{2}$ with $\alpha>0$ indicates the
normal attractive forces between the corresponding pair of nodes while 
$\beta\eta_{uv}^{2}$ with $\beta<0$ is transformed into an attractive term.

A case we haven't considered here is if $\alpha<0$
and $\beta>0$, when both terms represent repulsive forces between nodes. In 
this case $\alpha\xi_{uv}^{2}-\beta\eta_{uv}^{2}<0$
for all $u,v$ and the order in which the triads will be closed is
determined by the magnitudes of $\alpha$ and $\beta$.  In such a repulsive
system there are no attractive forces to fuel the creation of new edges. 
Consequently, the creation
of the new edges to close triads is controlled by  factors  such as their 
similarities or complementarity in their functions which
do not depend particularly on the communicability between nodes. In this 
case, predictions of triad closure made
on the basis of communicability distances are not expected to
differ significantly from those made by a random closure of the triads.

\begin{table}

\begin{centering}
\footnotesize
\begin{tabular}{cccccc}
\hline 
pair & $\xi_{uv}^{2}$  & $\eta_{uv}^{2}$ & $\xi_{uv}^{2}-1.5\eta_{uv}^{2}$  & $-0.5\xi_{uv}^{2}+1.5\eta_{uv}^{2}$ & $\xi_{uv}^{2}+\eta_{uv}^{2}$ \\
\hline 
{\footnotesize{}2,3 } & {\footnotesize{}2.000 } & {\footnotesize{}2.000 } & \textbf{\footnotesize{}-1.000}{\footnotesize{} } & {\footnotesize{}2.000 } & {\footnotesize{}4.000}\tabularnewline
{\footnotesize{}1,5 } & {\footnotesize{}3.184 } & {\footnotesize{}0.960 } & {\footnotesize{}1.744 } & \textbf{\footnotesize{}-0.152}{\footnotesize{} } & {\footnotesize{}4.144}\tabularnewline
{\footnotesize{}2,4 } & {\footnotesize{}2.545 } & {\footnotesize{}1.312 } & {\footnotesize{}0.577 } & {\footnotesize{}0.696 } & \textbf{\footnotesize{}3.857}\tabularnewline
\hline 
\end{tabular}
\par\end{centering}{\footnotesize \par}

{\footnotesize{}\protect\protect\caption{{\footnotesize{}Values of weighted sum of $\xi^{2}$ and $\eta^{2}$
for the potential edges considered in Figure 1.}}
}{\footnotesize \par}

{\footnotesize{}\label{Table of transformations} }
\end{table}

In summary, we can use the function 
\begin{equation}
\Delta_{uv}(\alpha,\beta):=\alpha\xi_{uv}^{2}-\beta\eta_{uv}^{2},\qquad\forall 
u,v\in V,\label{eq:delta}
\end{equation}
to determine which triad is closed in the network.

In order to predict which triads will close in a given network it
is necessary to know the values of $\alpha$
and $\beta$.  We now propose a method that allows
us to estimate these empirical parameters and consequently to determine
which triads will close in a given network.


\section{Proposed Method}

\label{sec:PM}

In order to predict the triadic closure in a network
 based on  $\Delta_{uv}(\alpha,\beta)$
we develop a procedure
to find the values of the empirical parameters $\alpha$
and $\beta$ which best predict the triadic closure in a network
from which we have\textit{ a priori }removed all the triangles. That
is, if we take a network $\Gamma$, we first detect all its existing
triangles. We then transform $\Gamma$ into a triangle-free
graph $\Gamma'$ by removing one and only one of the edges forming
each triangle. The deleted edges are selected uniformly at random from the
three edges forming each triangle. As this procedure is likely to be repeated a
large number of times (see below for details), the chance that
each of the three edges is selected at least once is very high. We keep a 
list of all these removed edges which we call $R$. It may happen that
two triangles $T_{1}$ and $T_{2}$ share an edge $e$. If we select
$e$ when considering $T_{1}$, then, when it comes to select an edge
in $T_{2}$, we pick an edge which may or may not coincide with $e$.
If it does, we do not add it to the list. It may also happen that
$T_{2}$ consists of $e$ and two other edges, one of which has also
already been removed because it was in common with a third triangle.
In such cases, we do not remove the last connection remaining in $T_{2}$,
since it could disconnect the network. 

We can also create a list, $P$, of all the pairs of nodes which
form triads in $\Gamma$ but were not part of any triangle.
Finally we create the list $L=R\cup P$. Because we have removed
one edge from each triangle, the nodes in $R$ are now separated by
two adjacent edges in $\Gamma'$, similarly to the
pairs of nodes in $P$. Our task is to select appropriate values of
the empirical parameters $\alpha$ and $\beta$ that differentiate
as much as possible the pairs of nodes in $R$ from those in $P$.
We do this by using a non-increasing ranking of all the pairs of nodes
in $L$ according to $\Delta_{uv}(\alpha,\beta)$. We have previously
predicted that the triadic closure process should be controlled by
the smallest values of $\Delta_{uv}(\alpha,\beta)$ (see example in
Figure \ref{transformation}). Thus, we expect that a non-increasing
ranking of the values of $\Delta_{uv}(\alpha,\beta)$ contains most
of the elements of $R$ at the top of the ranking and those of $P$
at the bottom. 

In order to quantify the percentage of triangles that were correctly
predicted we proceed as follow. We first rank the entries of $L$
in non-increasing order. We select the top $r$ entries of $L=R\cup P$,
where $r$ is the cardinality of $R$. Then, we count the number $r_{p}$
of entries in this top $r$ which are elements of $R$. These entries
correspond to those pairs of nodes which were originally part of the
triangles of $\Gamma$. That is, $r_{p}$ represents the number of
correct predictions made by the current method. We call the (percentage) ratio 
of $r_{p}$ to $r$  the percentage of \texttt{detected}.

\subsection{Datasets and Computational methods}

We now give some computational details on how
we implemented these calculations to find the optimal values of $\alpha$
and $\beta$ for a selection of networks.

We study 25 networks representing complex systems from a wide variety
of environments, such as social, ecological, biomolecular, technological,
infrastructural, and informational. A brief description of all these
networks is given in the Appendix. 

In order to find the optimal values of the empirical parameters $\alpha$
and $\beta$ for these networks we proceed as follows. We calculate
all the values of $\alpha$ and $\beta$ in the interval $I=[-2.1,2.1]$
with a step length of 0.1. This interval $I$ has been determined
empirically as smaller intervals led to worse results and larger ones
did not improve the results. Then, for each combination of $\alpha$
and $\beta$ in $\Delta_{uv}(\alpha,\beta)$ we rank all the elements
of $L$ in non-increasing order and find the percentage of {\tt detected}.
The optimal values of $\alpha$ and $\beta$ for this particular network
are those that produce the highest percentage of {\tt detected}.
These computations were repeated 100 times.

The effectiveness of the proposed method is tested by considering
a simple null model constructed as follows. We randomly order the
edges in $L$, select the top $r$ pairs of nodes and count how many
of them were in the set $R$. With this information we compute the
percentage of correct predictions made by a random ordering of the pairs
of nodes (\texttt{rand}). Similar values of the percentages of\texttt{
detected} and \texttt{rand} indicate that the ranking produced by
the function $\Delta_{uv}(\alpha,\beta)$ does not differ significantly
from a random ordering of the pairs of nodes and consequently is
not a good one; while larger differences between the percentages
of\texttt{ detected} and \texttt{rand }indicate good performance of
the proposed method.

Before starting with the detailed analysis of these 25 datasets we
consider the possibility of fixing one of the parameters ($\alpha$
or $\beta$) and letting the other varying in the bounded interval
$[-2.1,2.1]$. To do this we set $\alpha=1$ and let $\beta$ vary.
This seems reasonable, since this choice allows us to tune the disturbance
caused by the repulsion in the values of $\Delta_{uv}$. However,
the results obtained for 10 of the studied networks discouraged us
from proceeding with this approach. On average the use of the two
parameters $\alpha$ and $\beta$ makes predictions of triadic closure
which are $7\%$ higher than those using only one parameter, with
maximum differences of up to $20\%$ for one network (results not
shown here). Thus, we will use the more general approach of calibrating
both empirical parameters.

\subsection{Bounds for communicability distance functions}

Although in our experiments we use the exact values of the communicability
distance functions in order to obtain the values of $\Delta_{uv}$,
we now give some bounds for $\xi_{uv}$ and $\eta_{uv}$, which can be
used in the computations when working on extremely large networks.
It is clear from the definitions given in \eqref{eq:xi}, \eqref{eq:eta},
and \eqref{eq:delta} that for large matrices these values may be
too costly to compute. To avoid the computation of the matrix exponential,
we derive bounds for $\xi_{uv}^{2}$ and $\eta_{uv}^{2}$ (and therefore
for $\Delta_{uv}(\alpha,\beta)$) by means of a Gauss--Radau quadrature
rule. In order to make the present paper
self-contained, we summarize the approach used as described in 
\cite{Benzi2010,Benzi1999,Golub2010}  before giving these bounds.

It is well known that the problem of computing bilinear expressions
of the form $\mathbf{u}^{T}f(A)\mathbf{v}$ can be reduced to the
approximation of a Riemann--Stieltjes integral with respect to a certain
measure using quadrature rules. Indeed, in a series of papers, Golub
and collaborators use  1 step of
the symmetric Lanczos iteration to give bounds on the entries of $f(A)$ based 
on Gauss-type
quadrature rules when $f$ is a \textit{strictly completely monotonic}
(s.c.m.) function on an interval $\mathcal{I}$ containing the spectrum
of $A$. Recall that a function is 
s.c.m.\
on $\mathcal{I}$ if $f^{(2k)}(x)>0$ and $f^{(2k+1)}(x)<0$ for all
$x\in\mathcal{I}$ and $\forall k\geq0$. Since $g(x)=e^{x}$
is not s.c.m., we need to work on $f(x)=e^{-x}$ to derive bounds
on the quantities of interest here.

The key result that allows to easily compute \textit{a priori} bounds
using Gauss-type quadrature rules is that we can use
the element in position $(1,1)$ of the matrix $f(J_{p+1})$ (see
Theorem 6.6 in \cite{Golub2010}), where 
\[
J_{p+1}=\left(\begin{array}{ccccc}
\omega_{1} & \gamma_{1}\\
\gamma_{1} & \omega_{2} & \gamma_{2}\\
 & \ddots & \ddots & \ddots\\
 &  & \gamma_{p-1} & \omega_{p} & \gamma_{p}\\
 &  &  & \gamma_{p} & \omega_{p+1}
\end{array}\right)
\]
is a tridiagonal matrix whose eigenvalues are the nodes of the quadrature
rule, and the rule's weights are given by the squares of the first entries
of $J_{p+1}$'s normalized eigenvectors.

Our results are summarized in the following theorems. \begin{theorem}
\label{th:bound_cd} Let $A$ be the adjacency matrix of an unweighted
and undirected network. Then 
\begin{equation}
\Phi\left(b,\omega_{1}+\frac{\gamma_{1}^{2}}{\omega_{1}-b}\right)\leq\frac{
\left(\xi_{uv}\right)^{2}}{2}\leq\Phi\left(a,\omega_{1}+\frac{\gamma_{1}^{2}}{
\omega_{1}-a}\right),\label{eq:bound}
\end{equation}
where 
\begin{equation}
\Phi(x,y)=\frac{c\left(e^{-x}-e^{-y}\right)+xe^{-y}-ye^{-x}}{x-y},\qquad c=\omega_{1},\label{eq:Phi}
\end{equation}

\[
\left\{ \begin{array}{l}
\omega_{1}=a_{uv},\\
\gamma_{1}=\left[\frac{d_{u}+d_{v}}{2}-\omega_{1}-A_{uv}^{2}\right]^{\frac{1}{2}
},
\end{array}\right.
\]
and $[a,b]$ is an interval containing the spectrum of $-A$. \end{theorem}
\begin{rem} If $(u,v)\not\in E$ the bounds simplify considerably.
Indeed, in this case $\omega_{1}=0$ and hence 
\[
\frac{b^{2}e^{\frac{\gamma_{1}^{2}}{b}}+\gamma_{1}^{2}e^{-b}}{b^{2}+\gamma_{1}^{2}}\leq\frac{\left(\xi_{uv}\right)^{2}}{2}\leq\frac{a^{2}e^{\frac{\gamma_{1}^{2}}{a}}+\gamma_{1}^{2}e^{-a}}{a^{2}+\gamma_{1}^{2}}
\]
 \end{rem}

Before proceeding with the proof of the result, note that $(\xi_{uv})^{2}$
can be written as 
\[
\left(\xi_{uv}\right)^{2}=(\mathbf{e}_{u}-\mathbf{e}_{v})^{T}\left(e^{A}\right)(\mathbf{e}_{u}-\mathbf{e}_{v}),
\]
where $\mathbf{e}_{u}$ and $\mathbf{e}_{v}$ are the $u$th and $v$th
vectors of the canonical basis, respectively. \begin{proof} Using
the Lagrange interpolation formula for the evaluation of matrix functions
\cite{FunctionMAtrixBook} one can easily show \cite{Benzi1999}
that 
\[
\mathbf{e}_{1}^{T}(e^{-C})\mathbf{e}_{1}=\frac{c_{11}(e^{-\mu_{1}}-e^{-\mu_{2}})+\mu_{1}e^{-\mu_{2}}-\mu_{2}e^{\mu_{1}}}{\mu_{1}-\mu_{2}}.
\]
where $\mu_{1}$, $\mu_{2}$ are the distinct eigenvalues of the matrix
$C$.

We now want to build explicitly the matrix $J_{2}=\left(\begin{array}{cc}
\omega_{1} & \gamma_{1}\\
\gamma_{1} & \omega_{2}
\end{array}\right)$ in such a way that $\tau_{1}=a$ or $\tau_{1}=b$ is a prescribed
eigenvalue. The values of $\omega_{1}$ and $\gamma_{1}$ are derived
explicitly by applying one step of Lanczos iteration to the matrix $-A$
with starting vectors $\mathbf{x}_{-1}=\mathbf{0}$ and 
$\mathbf{x}_{0}=(\mathbf{e}_{u}-\mathbf{e}_{v})/{\sqrt{2}}$.

Note that if $\gamma_{1}=0$ we simply take $\omega_{2}=\tau_{1}$
and the matrix $J_{2}$ is diagonal with eigenvalues $\mu_{1}=\omega_{1}$
and $\mu_{2}=\tau_{1}$. Thus, let us assume $\gamma_{1}\neq0$. Using
the three-term recurrence for orthogonal polynomials: 
\[
\gamma_{j}p_{j}(\lambda)=(\lambda-\omega_{j})p_{j-1}(\lambda)-\gamma_{j-1}p_{j-2}(\lambda),\quad j=1,2,\ldots,p,
\]
with $p_{-1}(\lambda)\equiv0$, $p_{0}(\lambda)\equiv1$ we find that
$\omega_{2}=\tau_{1}-\frac{\gamma_{1}}{p_{1}(\tau_{1})}$. Using the
same recurrence, we also find that 
$p_{1}(\tau_{1})=({\tau_{1}-\omega_{1}})/{\gamma_{1}}\neq0$,
since the zeros of orthogonal polynomials satisfying the three-term
recurrence are distinct and lie in the interior of $\mathcal{I}$
(see \cite[Theorem 2.14]{Golub2010}).

The matrix 
\[
J_{2}=\left(\begin{array}{cc}
\omega_{1} & \gamma_{1}\\
\gamma_{1} & \tau_{1}-\frac{\gamma_{1}^{2}}{\tau_{1}-\omega_{1}}
\end{array}\right)
\]
has (distinct) eigenvalues $\mu_{1}=\tau_{1}$ and $\mu_{2}=\omega_{1}+\frac{\gamma_{1}^{2}}{\omega_{1}-\tau_{1}}$.
The result then follows by applying Theorems 6.4 and 6.6 from \cite{Golub2010}.
\qquad{} \end{proof}

Similar bounds can be computed for $\eta_{uv}^{2}$
and are described in the following theorem, whose proof matches that
of Theorem \ref{th:bound_cd}. \begin{theorem} Let $A$ be the adjacency
matrix of an unweighted and undirected network. Then 
\[
\Phi\left(\tilde{b},\tilde{\omega}_{1}+\frac{\tilde{\gamma}_{1}^{2}}{\tilde{\omega}_{1}-\tilde{b}}\right)\leq\frac{(\eta_{uv})^{2}}{2}\leq\Phi\left(\tilde{a},\tilde{\omega}_{1}+\frac{\tilde{\gamma}_{1}^{2}}{\tilde{\omega}_{1}-\tilde{a}}\right)
\]
where $\Phi$ is defined as in equation (\ref{eq:Phi}) with $c=\tilde{\omega}_{1}$,
$\tilde{\mathcal{I}}=[\tilde{a},\tilde{b}]$ is an interval containing
the spectrum of $A^{2}$, and 
\[
\left\{ \begin{array}{l}
\tilde{\omega}_{1}=\gamma_{1}^{2}+\omega_{1};\\
\tilde{\gamma}_{1}=\left[\frac{1}{2}\sum_{w=1}^{n}\left(A_{uw}^{2}-A_{wv}^{2}\right)^{2}-\tilde{\omega_{1}}^{2}\right]^{\frac{1}{2}}
\end{array}\right..
\]
with $\omega_{1}$ and $\gamma_{1}$ as in theorem \ref{th:bound_cd}.
\end{theorem}

\begin{rem} Since the behavior of the eigenvalues of $A$ is known
(see \cite{vanMieghemBook}), we may take $\tilde{a}=0$ and $\tilde{b}=a^{2}$
as the square of the approximation to the largest eigenvalue of $A$.
For these values, the bounds simplify to
\[
\Phi\left(a^{2},\tilde{\omega}_{1}+\frac{\tilde{\gamma}_{1}^{2}}{\tilde{\omega}_{1}-a^{2}}\right)\leq\frac{(\eta_{uv})^{2}}{2}\leq\Phi\left(0,\tilde{\omega}_{1}+\frac{\tilde{\gamma}_{1}^{2}}{\tilde{\omega}_{1}}\right)=\frac{\tilde{\omega}_{1}^{2}e^{-\frac{\tilde{\omega}_{1}^{2}+\tilde{\gamma}_{1}^{2}}{\tilde{\omega}_{1}}}+\tilde{\gamma}_{1}^{2}}{\tilde{\omega}_{1}^{2}+\tilde{\gamma}_{1}^{2}}.
\]
 \end{rem}

Combining the results described in the previous theorems, one easily
get bounds for the values of $\frac{\Delta_{uv}(\gamma,\beta)}{2}$.
Indeed, the computation is straightforward, since the new bounds are
linear combinations of the previous ones. For example,
if we have $\gamma\geq0$ and $\beta\leq0$ we get as lower bound
for $\frac{\Delta_{uv}(\gamma,\beta)}{2}$ 
\[
\gamma\Phi\left(b,\omega_{1}+\frac{\gamma_{1}^{2}}{\omega_{1}-b}
\right)+\beta\Phi\left(\tilde{a},\tilde{\omega}_{1}+\frac{\tilde{\gamma}_{1}^{2}
}{\tilde{\omega}_{1}-\tilde{a}}\right),
\]
and as upper bound 
\[
\gamma\Phi\left(a,\omega_{1}+\frac{\gamma_{1}^{2}}{\omega_{1}-a}
\right)+\beta\Phi\left(\tilde{b},\tilde{\omega}_{1}+\frac{\tilde{\gamma}_{1}^{2}
}{\tilde{\omega}_{1}-\tilde{b}}\right),
\]
where $\omega_{1}$, $\gamma_{1}$, $\tilde{\omega}_{1}$, and $\tilde{\gamma}_{1}$
depend on the choice of $u$ and $v$.


\section{Modeling Results and Discussion}

\label{sec:MRD}

\subsection{Predicting and interpreting triadic closure}

The first series of results refers to the finding of the optimal values
of $\alpha$ and $\beta$ for the studied networks and the comparison
of the percentage of triadic closures correctly predicted by the proposed
method in comparison with the random process. The results of the tests
are listed in Table \ref{tab:2param_-A2R}. The columns $\langle\alpha^{*}\rangle$
and $\langle\beta^{*}\rangle$ contain the average best values for
the parameters, where the average is taken over the $100$ iterations
we run. The results show that on average the method based on the communicability
distance functions (\texttt{detect}) correctly predicts $20\%$ of
the triad closures in the real-world networks studied. In 7 cases
this percentage of correct prediction reaches values larger than $25\%$.
In contrast, the random closure of triads identifies $7.6\%$ of the
real triangles existing in those networks.

We can now gain some insights about the processes that have governed
the triad closure in the studied networks. Recall that in general
the triadic closure process consists of two different means of transmission
of information, namely the TEC and the LRC. If we refer to the nature
of the two kinds of transmission in the order TEC-LRC we can have
any of the following four scenarios: 
\begin{itemize}
\item $\alpha>0$, $\beta<0$, the triads close by means of attractive-attractive
interactions; 
\item $\alpha>0$, $\beta>0$, the triads close by means of attractive-repulsive
interactions; 
\item $\alpha<0$, $\beta<0$, the triads close by means of repulsive-attractive
interactions; 
\item $\alpha<0$, $\beta>0$, the triads close by means of repulsive-repulsive
interactions. 
\end{itemize}
In Table \ref{tab:2param_-A2R} we have arranged the values of $\langle\alpha^{*}\rangle$
and $\langle\beta^{*}\rangle$ to correspond to these four
classes. For instance, the networks Sawmill, Social3, Matheoremethod,
Galesburg, Prison, Zachary, and Colorado (all social networks), Grassland
and Bridge Brook (food webs) and Transc\_yeast (a gene transcription
network) close their triads following a scheme of attractive-attractive
interactions. The three social networks of High Tech, Drugs and Geom
as well as the networks of USAir97 (air transportation network), neurons
(neural network), Ythan1 (a food web) and the Internet at Autonomous
System level, all belong to the class of networks in which triads are closed
by an attraction-repulsion mechanism. The only network with a repulsion-attraction
triad closure mechanism is the social network of High School, while
there are 7 networks closing triads with a repulsion-repulsion mechanism
(three protein-protein interaction (PPIs) networks, two food webs, one animal social network and the Roget
thesaurus).

The group of networks with attractive-attractive interactions consists
of $63\%$ of all the social networks studied here. Among them we
find a communication network within a small enterprise: a sawmill,
where all employees were asked to indicate the frequency with which
they discussed work matters with each of their colleagues on a five-point
scale ranging from less than once a week to several times a day. Two
employees were linked in the communication network if they rated their
contact as a three or higher. This is a cooperative process
in which both advisers and advisees cooperate to share the information
needed for improving their skills and knowledge. Thus, closing the
potential triangles in order to enhance the communication between
the individuals involved seems a very reasonable mechanism. The other
social networks included in this class all share a common characteristic.
In the networks Social3 (a network of social contacts among college
students participating in a leadership course), Galesburg (a network
of friendship among physicians) and Matheoremethod (a network of friendship
among school superintendents) the participants in the respective studies
were asked the following questions: 
\begin{itemize}
\item Choose the three members they wished to include in a committee; 
\item Nominate three doctors with whom they would choose to discuss medical
matters; 
\item Name their best friends among the chief school administrators in Allegheny
County. 
\end{itemize}
In the first two cases it is very clear that the participants were
asked to nominate individuals with whom they would easily cooperate,
e.g., members of a committee or colleagues with whom to discuss medical
matters. The third resembles a general kind of friendship
relation, but the question was formulated in the context of analyzing
the diffusion of a new mathematical method among High Schools in the
county. Thus, selecting your best friends among other chief school
administrators also means selecting those with whom you would easily
cooperate in technical matters. These facts may explain the kind of
attraction-attraction interaction which dictates the main mechanism
for closing the triads in these networks. Transmission of information
through the edges as well as a direct long-range interaction between
peers both benefit the cooperative atmosphere needed for performing
the tasks for which these networks are created.

In the class of networks in which triads have been closed by attraction-repulsion
mechanisms we find networks of very different natures and it is difficult
to extend the previous analysis to all these networks. This
group includes a social network in a small high-tech computer firm
which sells, installs, and maintains computer systems, where individuals were 
asked:
``\textit{Who do you consider to be a personal friend?}''. It could
be speculated that a mechanism of the type based on Simmelian principles
dominates here. That is, if $A-B-C$ is a triad and the two pairs
$A-B$ and $B-C$ have strong social relations, it is natural to think
that there is not a strong repulsion between $A$ and $C$ and they can 
create a new social tie. The friendship network among boys in a High School,
which is the only one showing repulsion-attraction mechanisms, was created
by asking the pupils: ``\textit{What fellows here in school do you
go around with most often?}''. The triads here are formed when the
relations between the pairs $A-B$ and $B-C$ are not strong enough
to tie $A$ and $C$ together. If the pairs $A-B$ and $B-C$
have some strong relation, i.e. if they are dating, a link between
$A$ and $C$ could be seen as offensive to the
already established couples. The final class of networks, that characterized
by repulsion-repulsion interactions, does not contain any human social
network. The three PPIs included in this study are characterized by
this type of triad closure mechanism, together with 2 food webs, an
animal social network and a thesaurus. The repulsion-repulsion mechanism
is characterized by weak through-edge transmission of information
and weak long-range interaction between pairs of nodes separated by
two adjacent edges. Thus, it is expected that the triad closure is
controlled by non-topological factors, such as similarities or complimentarities
among the nodes. This is a plausible explanation for the case of the
PPI networks where triads of proteins may form triangles due to their
functional similarities. We notice that, as expected, the percentages
of correct prediction of triad closure in this group are the smallest
of the four groups. That is, the difference between the predictions
made by the proposed method and the random one in this group is 8.7\%,
in contrasts with 15.5\% for the attraction-attraction, 14.1\% for
the attraction-repulsion and 10\% for the only network with repulsion-attraction
mechanisms.

\begin{table}
\centering 
\footnotesize
\caption{Results of the proposed method for predicting triad
closure in 25 complex networks.}
\begin{tabular}{lccccc}
\hline 
{\footnotesize{}{}Network  } & {\footnotesize{}{}r  } & \texttt{\footnotesize{}{}detected}{\footnotesize{}{}  } & \texttt{\footnotesize{}{}rand}{\footnotesize{}{}  } & {\footnotesize{}{}$\langle\alpha^{*}\rangle$  } & {\footnotesize{}{}$\langle\beta^{*}\rangle$ }\tabularnewline
\hline 
{\footnotesize{}{}{}{}Sawmill  } & {\footnotesize{}{}{}{}18  } & {\footnotesize{}{}{}{}$27\%$  } & {\footnotesize{}{}{}{}$10\%$  } & {\footnotesize{}{}{}{}$1.906$  } & {\footnotesize{}{}{}{}$-1.25$ }\tabularnewline
{\footnotesize{}{}{}{}social3  } & {\footnotesize{}{}{}{}32  } & {\footnotesize{}{}{}{}$24\%$  } & {\footnotesize{}{}{}{}$11\%$  } & {\footnotesize{}{}{}{}$1.164$  } & {\footnotesize{}{}{}{}$-1.258$ }\tabularnewline
{\footnotesize{}{}{}{}Matheoremethod  } & {\footnotesize{}{}{}{}19  } & {\footnotesize{}{}{}{}$25\%$  } & {\footnotesize{}{}{}{}$10\%$  } & {\footnotesize{}{}{}{}$1.196$  } & {\footnotesize{}{}{}{}$-0.574$ }\tabularnewline
{\footnotesize{}{}{}{}Grassland  } & {\footnotesize{}{}{}{}30  } & {\footnotesize{}{}{}{}$25\%$  } & {\footnotesize{}{}{}{}$9\%$  } & {\footnotesize{}{}{}{}$1.833$  } & {\footnotesize{}{}{}{}$-1.203$ }\tabularnewline
{\footnotesize{}{}{}{}Galesburg  } & {\footnotesize{}{}{}{}29  } & {\footnotesize{}{}{}{}$23\%$  } & {\footnotesize{}{}{}{}$11\%$  } & {\footnotesize{}{}{}{}$0.902$  } & {\footnotesize{}{}{}{}$-0.648$ }\tabularnewline
{\footnotesize{}{}{}{}Prison  } & {\footnotesize{}{}{}{}58  } & {\footnotesize{}{}{}{}$30\%$  } & {\footnotesize{}{}{}{}$12\%$  } & {\footnotesize{}{}{}{}$0.294$  } & {\footnotesize{}{}{}{}$-1.492$ }\tabularnewline
{\footnotesize{}{}{}{}Zachary  } & {\footnotesize{}{}{}{}45  } & {\footnotesize{}{}{}{}$42\%$  } & {\footnotesize{}{}{}{}$10\%$  } & {\footnotesize{}{}{}{}$1.696$  } & {\footnotesize{}{}{}{}$-0.392$ }\tabularnewline
{\footnotesize{}{}{}{}BridgeBrook  } & {\footnotesize{}{}{}{}774  } & {\footnotesize{}{}{}{}$13\%$  } & {\footnotesize{}{}{}{}$3\%$  } & {\footnotesize{}{}{}{}$1.977$  } & {\footnotesize{}{}{}{}$-1.046$ }\tabularnewline
{\footnotesize{}{}{}{}Colorado  } & {\footnotesize{}{}{}{}17  } & {\footnotesize{}{}{}{}$20\%$  } & {\footnotesize{}{}{}{}$1\%$  } & {\footnotesize{}{}{}{}$0.754$  } & {\footnotesize{}{}{}{}$-0.044$ }\tabularnewline
{\footnotesize{}{}{}{}Transc\_yeast  } & {\footnotesize{}{}{}{}72  } & {\footnotesize{}{}{}{}$4\%$  } & {\footnotesize{}{}{}{}$1\%$  } & {\footnotesize{}{}{}{}$0.221$  } & {\footnotesize{}{}{}{}$-0.544$ }\tabularnewline
\hline 
{\footnotesize{}{}{}{}USAir97  } & {\footnotesize{}{}{}{}12181  } & {\footnotesize{}{}{}{}$45\%$  } & {\footnotesize{}{}{}{}$18\%$  } & {\footnotesize{}{}{}{}$1.452$  } & {\footnotesize{}{}{}{}$0.63$ }\tabularnewline
{\footnotesize{}{}{}{}High tech  } & {\footnotesize{}{}{}{}77  } & {\footnotesize{}{}{}{}$31\%$  } & {\footnotesize{}{}{}{}$16\%$  } & {\footnotesize{}{}{}{}$0.198$  } & {\footnotesize{}{}{}{}$0.288$ }\tabularnewline
{\footnotesize{}{}{}{}Drugs  } & {\footnotesize{}{}{}{}3598  } & {\footnotesize{}{}{}{}$27\%$  } & {\footnotesize{}{}{}{}$16\%$  } & {\footnotesize{}{}{}{}$0.526$  } & {\footnotesize{}{}{}{}$1.048$ }\tabularnewline
{\footnotesize{}{}{}{}Neurons  } & {\footnotesize{}{}{}{}2808  } & {\footnotesize{}{}{}{}$16\%$  } & {\footnotesize{}{}{}{}$8\%$  } & {\footnotesize{}{}{}{}$0.526$  } & {\footnotesize{}{}{}{}$0.978$ }\tabularnewline
{\footnotesize{}{}{}{}Geom  } & {\footnotesize{}{}{}{}12325  } & {\footnotesize{}{}{}{}$12\%$  } & {\footnotesize{}{}{}{}$6\%$  } & {\footnotesize{}{}{}{}$0.14$  } & {\footnotesize{}{}{}{}$1.149$ }\tabularnewline
{\footnotesize{}{}{}{}Ythan1  } & {\footnotesize{}{}{}{}507  } & {\footnotesize{}{}{}{}$10\%$  } & {\footnotesize{}{}{}{}$4\%$  } & {\footnotesize{}{}{}{}$0.248$  } & {\footnotesize{}{}{}{}$0.492$ }\tabularnewline
{\footnotesize{}{}{}{}Internet  } & {\footnotesize{}{}{}{}2331  } & {\footnotesize{}{}{}{}$26\%$  } & {\footnotesize{}{}{}{}$0\%$  } & {\footnotesize{}{}{}{}$0.1$  } & {\footnotesize{}{}{}{}$1.842$ }\tabularnewline
\hline 
{\footnotesize{}{}{}{}High School  } & {\footnotesize{}{}{}{}199  } & {\footnotesize{}{}{}{}$28\%$  } & {\footnotesize{}{}{}{}$18\%$  } & {\footnotesize{}{}{}{}$-0.654$  } & {\footnotesize{}{}{}{}$-0.434$ }\tabularnewline
\hline 
{\footnotesize{}{}{}{}Dolphins  } & {\footnotesize{}{}{}{}95  } & {\footnotesize{}{}{}{}$24\%$  } & {\footnotesize{}{}{}{}$13\%$  } & {\footnotesize{}{}{}{}$-0.364$  } & {\footnotesize{}{}{}{}$0.586$ }\tabularnewline
{\footnotesize{}{}{}{}ScotchBroom  } & {\footnotesize{}{}{}{}358  } & {\footnotesize{}{}{}{}$31\%$  } & {\footnotesize{}{}{}{}$4\%$  } & {\footnotesize{}{}{}{}$-0.372$  } & {\footnotesize{}{}{}{}$0.660$ }\tabularnewline
{\footnotesize{}{}{}{}StMartin  } & {\footnotesize{}{}{}{}278  } & {\footnotesize{}{}{}{}$16\%$  } & {\footnotesize{}{}{}{}$11\%$  } & {\footnotesize{}{}{}{}$-0.232$  } & {\footnotesize{}{}{}{}$0.335$ }\tabularnewline
{\footnotesize{}{}{}{}PIN\_Ecoli  } & {\footnotesize{}{}{}{}478  } & {\footnotesize{}{}{}{}$10\%$  } & {\footnotesize{}{}{}{}$5\%$  } & {\footnotesize{}{}{}{}$-1.025$  } & {\footnotesize{}{}{}{}$0.137$ }\tabularnewline
{\footnotesize{}{}{}{}PIN\_Yeast  } & {\footnotesize{}{}{}{}3530  } & {\footnotesize{}{}{}{}$13\%$  } & {\footnotesize{}{}{}{}$4\%$  } & {\footnotesize{}{}{}{}$-1.53$  } & {\footnotesize{}{}{}{}$1.842$ }\tabularnewline
{\footnotesize{}{}{}{}PIN\_Human  } & {\footnotesize{}{}{}{}1047  } & {\footnotesize{}{}{}{}$5\%$  } & {\footnotesize{}{}{}{}$2\%$  } & {\footnotesize{}{}{}{}$-0.203$  } & {\footnotesize{}{}{}{}$0.291$ }\tabularnewline
{\footnotesize{}{}{}{}Roget  } & {\footnotesize{}{}{}{}1550  } & {\footnotesize{}{}{}{}$7\%$  } & {\footnotesize{}{}{}{}$6\%$  } & {\footnotesize{}{}{}{}$-0.305$  } & {\footnotesize{}{}{}{}$0.008$ }\tabularnewline
\hline 
\end{tabular}{\footnotesize{}{}{}{}\label{tab:2param_-A2R}  }
\end{table}

\subsection{Network evolution under triadic closure}

Finally in this section we use the results of the proposed method
for modeling the triadic closure evolution in a given network. Although
we can model the future evolution of a network from its current state,
we prefer to consider a network in an early stage of its evolution
and to predict how it has evolved towards its current structure. This
method allows us to contrast the predictions made by the current method
with some control parameters obtained for the real-world network in
its current state. For this experiment we selected the network of
injecting drug users (Drugs) for which we consider the clustering
coefficient, the average path length, and the average communicability
of the actual network. In order to perform these experiments we select
$50\%$ of the total number of triangles existing in the network and
we remove one edge from each of them. Edges are selected uniformly
at random among those present in the corresponding triangle. As before,
let $L$ be the list of edges obtained from the union of the potential
edges and of those we removed. The values for $\alpha^{*}$ and $\beta^{*}$
are those determined empirically using the calibration method already
described (cf. Table \ref{tab:2param_-A2R}).

The iteration process goes as follows. We select the potential edge
realizing the smallest value for $\Delta_{uv}(\alpha^{*},\beta^{*})$
and we add this edge to the network. Then we compute the values of
the parameters of interest: the average clustering coefficient, the
average path length, and the average communicability. Finally, the
values for $\Delta_{uv}(\alpha^{*},\beta^{*})$ are recomputed using
the new adjacency matrix, obtained by the
addition of the selected potential edge. Here every addition of an
edge is considered as a time step.

This iteration is run as many times as the number of edges we have
removed. That is, if we removed $r$ edges, we consider a discrete
time evolution for $0\leq t\leq r$. We then repeat this experiment
10 times, taking the average and standard deviations of the corresponding
parameter. In order to compare the results we simulate the same process
by adding such edges uniformly at random.

The results of this experiment are illustrated in Figure 2, where
we plot the values for the parameters of interest (with the corresponding
error bar) versus time. The horizontal dotted line represents the
actual value of the property for the original real-world network.
As can be seen in Figure 2, the proposed method outperforms the
random one for predicting the clustering coefficient of the network.
The current value of $\overline{C}$ for this network is $0.549$,
while the one predicted by $\Delta_{uv}$ is $0.486$, which contrasts
with that of $0.183$ obtained by the random method. We remark here
that this is the most important parameter to be considered in this
experiment as it is the one which accounts more directly for the ratio
of triangles to paths of length two in the network. Both methods predict
the average path length of the network very well, returning values
very close to the actual one ($\overline{\ell}=5.287$). In addition,
the proposed method increases  the average communicability
of the network more significantly than the random triadic closure. 
This feature
is important when one is interested in maximizing the total
average communicability of a network, which is equivalent to increasing
the quality of communication among the nodes in the network.

\begin{figure}
\centering \includegraphics[width=0.9\textwidth]{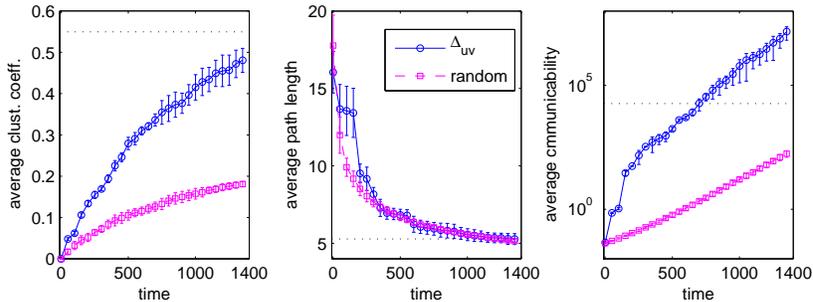}
\label{fig:evolution10} \protect\caption{Illustration of the evolution of the clustering coefficient, average
path length, and average communicability for the network of injecting
drug users (Drugs) versus the number of links added using the function
($\Delta_{uv}$) and at random (see the text for explanations).}
\end{figure}

\section{Conclusions}

\label{sec:conclusions}

The prediction of triadic closure is a very important and far from
trivial problem in network theory. The fact that most   real-world
complex networks have more triangles than random counterparts makes
the problem interesting \textit{per se}. In addition, there is a large
amount of evidence that shows that triadic closure in (social)
networks is an important driver for other important structural characteristics
of networks, such as degree distributions, clustering, and community
structure. In this work, we introduce a triad closure mechanism based
on communicability distances among pairs of nodes in a network. Our
results show acceptable levels of predictability and interpretability
of the potential mechanisms controlling triad closure in real-world
networks.

\section*{Acknowledgements}

E.E. thanks the Royal Society for a Wolfson Research Merit Award.


\Appendix
\section*{}

In this section we give a brief description of the networks used for
the tests throughout the paper.

\begin{table}[htb]
\centering
\footnotesize
\begin{tabular}{cccc}
\hline 
Name  & $n$  & {\footnotesize{}$r$ } & {\footnotesize{}$|P|$ }\\
\hline 
Matheoremethod  & {\footnotesize{}30 } & {\footnotesize{}19 } & {\footnotesize{}175}\tabularnewline
Galesburg  & {\footnotesize{}31 } & {\footnotesize{}29 } & {\footnotesize{}224 }\tabularnewline
High tech  & {\footnotesize{}33 } & {\footnotesize{}77 } & {\footnotesize{}390 }\tabularnewline
Zachary  & {\footnotesize{}34 } & {\footnotesize{}45 } & {\footnotesize{}393 }\tabularnewline
Sawmill  & {\footnotesize{}36 } & {\footnotesize{}18 } & {\footnotesize{}165 }\tabularnewline
social3  & {\footnotesize{}37 } & {\footnotesize{}32 } & {\footnotesize{}299 }\tabularnewline
StMartin  & {\footnotesize{}44 } & {\footnotesize{}278 } & {\footnotesize{}1732 }\tabularnewline
Dolphins  & {\footnotesize{}62 } & {\footnotesize{}95 } & {\footnotesize{}638 }\tabularnewline
Prison & {\footnotesize{}67 } & {\footnotesize{}58 } & {\footnotesize{}430 }\tabularnewline
High School  & {\footnotesize{}69 } & {\footnotesize{}199 } & {\footnotesize{}874 }\tabularnewline
BridgeBrook & {\footnotesize{}75 } & {\footnotesize{}774 } & {\footnotesize{}9829 }\tabularnewline
Grassland  & {\footnotesize{}75 } & {\footnotesize{}30 } & {\footnotesize{}427 }\tabularnewline
Ythan1  & {\footnotesize{}134 } & {\footnotesize{}507 } & {\footnotesize{}9019 }\tabularnewline
ScotchBroom  & {\footnotesize{}154 } & {\footnotesize{}358 } & {\footnotesize{}4094 }\tabularnewline
PIN\_Ecoli  & {\footnotesize{}230 } & {\footnotesize{}478 } & {\footnotesize{}7803 }\tabularnewline
Neurons & {\footnotesize{}280 } & {\footnotesize{}2808 } & {\footnotesize{}33973 }\tabularnewline
USAir97  & {\footnotesize{}332 } & {\footnotesize{}12181 } & {\footnotesize{}55646 }\tabularnewline
Colorado & {\footnotesize{}324 } & {\footnotesize{}17 } & {\footnotesize{}1273 }\tabularnewline
Drugs  & {\footnotesize{}616 } & {\footnotesize{}3598 } & {\footnotesize{}18533 }\tabularnewline
Transc\_yeast  & {\footnotesize{}662 } & {\footnotesize{}72 } & {\footnotesize{}13069 }\tabularnewline
Roget  & {\footnotesize{}994 } & {\footnotesize{}1550 } & {\footnotesize{}30116 }\tabularnewline
PIN\_Yeast & {\footnotesize{}2224 } & {\footnotesize{}3530 } & {\footnotesize{}92882 }\tabularnewline
PIN\_Human  & {\footnotesize{}2783 } & {\footnotesize{}1047 } & {\footnotesize{}85617 }\tabularnewline
Internet  & {\footnotesize{}3015 } & {\footnotesize{}2331 } & {\footnotesize{}462232}\tabularnewline
Geom  & {\footnotesize{}3621 } & {\footnotesize{}12325 } & {\footnotesize{}127794 }\tabularnewline
\hline 
\end{tabular}
\caption{Dataset: $n$ number of nodes in the network, $r$
number of existing triangles, and $|P|$ number of open triads. }
\label{tab:dataset} 
\end{table}

\textit{Brain networks} 
\begin{itemize}
\item Neurons: Neuronal synaptic network of the nematode \textit{C.~elegans}.
Includes all data except muscle cells and uses all synaptic connections
\cite{White1986,Milo2002}. 
\end{itemize}
\textit{Ecological networks} 
\begin{itemize}
\item BridgeBrook: pelagic species from the largest of a set of 50 New York
Adirondack lake food webs \cite{Polis1991}; 
\item Grassland: all vascular plants and all insects and trophic interactions
found inside stems of plants collected from 24 sites distributed within
England and Wales \cite{Martinez1999}; 
\item ScotchBroom: trophic interactions between the herbivores, parasitoids,
predators and pathogens associated with broom,\textit{Cytisus scoparius},
collected in Silwood Park, Berkshire, England, UK \cite{Memmott2000}; 
\item StMartin: birds and predators and arthropod prey of Anolis lizards
on the island of St. Martin, which is located in the northern Lesser
Antilles \cite{Martinez1991}; 
\item Ythan1: mostly birds, fishes, invertebrates, and metazoan parasites
in a Scottish Estuary \cite{Huxman1996}. 
\end{itemize}
\textit{Informational networks} 
\begin{itemize}
\item Roget: vocabulary network of words related by their definitions in
Roget Thesaurus of English. Two words are connected if one is used
in the definition of the other \cite{Roget2002}. 
\end{itemize}
\textit{PPI networks} 
\begin{itemize}
\item PIN\_Ecoli: protein-protein interaction network in \textit{Escherichia
coli}\cite{Butland2005}; 
\item PIN\_Human: protein-protein interaction network in human \cite{Rual2005}; 
\item PIN\_Yeast: protein-protein interaction network in \textit{S.~cerevisiae}
(yeast) \cite{Bu2003,vonMering2002}. 
\end{itemize}
\textit{Social and economic networks} 
\begin{itemize}
\item Colorado: the risk network of persons with HIV infection during its
early epidemic phase in Colorado Spring, USA, using analysis of community
wide HIV/AIDS contact tracing records (sexual and injecting drugs
partners) from 1985-1999 \cite{Potterat2002}; 
\item Dolphins: social network of frequent association between $62$ bottlenose
dolphins living in the waters off New Zealand \cite{Lusseau2003}; 
\item Drugs: social network of injecting drug users (IDUs) that have shared
a needle in the last six months \cite{Moody2001}. 
\item Galesburg: friendship ties among 31 physicians \cite{Coleman1966,Knoke1983,deNooy2005}; 
\item Geom: collaboration network of scientists in the field of Computational
Geometry \cite{Batagelj2006}; 
\item High School: network of relations in a high school. The students choose
the three members they wanted to have in a committee \cite{Zeleny1950}; 
\item High tech: friendship ties among the employees in a small high-tech
computer firm which sells, installs, and maintain computer systems
\cite{Krackhardt1999,deNooy2005}; 
\item Matheoremethod: this network concerns the diffusion of a new mathematics
method in the 1950s. It traces the diffusion of the modern mathematical
method among school systems that combine elementary and secondary
programs in Allegheny County (Pennsylvania, U.S.) \cite{Carlson1965,deNooy2005}; 
\item Prison: social network of inmates in prison who chose ``What fellows
on the tier are you closest friends with?'' \cite{MacRae1960}; 
\item Sawmill: social communication network within a sawmill, where employees
were asked to indicate the frequency with which they discussed work
matters with each of their colleagues \cite{Michael1997,deNooy2005}; 
\item social3: social network among college students in a course about leadership.
The students choose which three members they wanted to have in a committee
\cite{Zeleny1950}; 
\item Zachary: social network of friendship among the members of a karate
club \cite{Zachary1977}. 
\end{itemize}
\textit{Technological networks} 
\begin{itemize}
\item Internet: the Internet at the Autonomous System (AS) level as of September
1997 and of April 1998 \cite{Faloutsos1999}; 
\item USAir97: airport transportation network between airports in US in
1997 \cite{Batagelj2006}. 
\end{itemize}
\textit{Transcription networks} 
\begin{itemize}
\item Transc\_yeast: direct transcriptional regulation between genes in
\textit{Saccaromyces cerevisiae} \cite{Milo2002,Milo2004}. 
\end{itemize}




\end{document}